\documentclass[10pt, conference, compsocconf]{IEEEtran}

\usepackage{cite}

\usepackage{graphicx}

\usepackage{amssymb}
\usepackage{amsmath}

\usepackage{multirow}

\usepackage{subfigure}

\usepackage[hyphens]{url}
\usepackage{hyperref}

\usepackage{lipsum}

\usepackage{pdfpages}

\newcommand\blfootnote[1]{%
  \begingroup
  \renewcommand\thefootnote{}\footnote{#1}%
  \addtocounter{footnote}{-1}%
  \endgroup
}

\begin{document}

\includepdf[pages={1}]{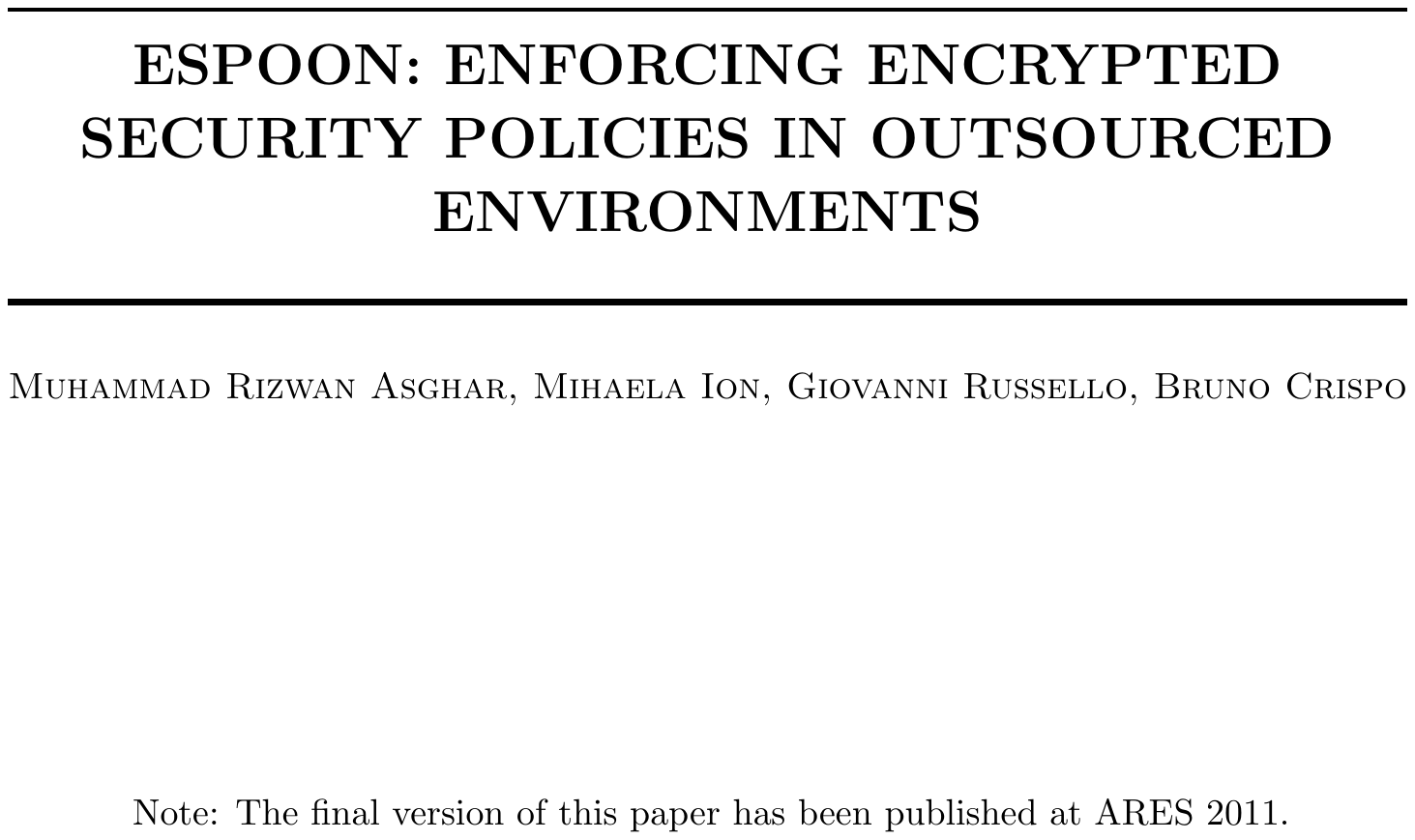}

\mathchardef\mhyphen="2D

\pagestyle{plain}
\pagenumbering{arabic}


\title{ESPOON: Enforcing Encrypted Security Policies in Outsourced Environments}

\author{
\IEEEauthorblockN{Muhammad Rizwan Asghar, Mihaela Ion,\\ Giovanni Russello}
\IEEEauthorblockA{Create-Net\\
Trento, Italy\\
\{asghar, ion, russello\}@create-net.org}
\and
\IEEEauthorblockN{Bruno Crispo}
\IEEEauthorblockA{Department of \\ Information Engineering and Computer Science\\
University of Trento, Italy\\
crispo@disi.unitn.it}
}

\maketitle

\blfootnote{* The final version of this paper has been published at ARES 2011 \cite{Asghar2011-ARES}.}

\begin{abstract}
The enforcement of security policies in outsourced environments is still an open challenge for policy-based systems. On the one hand, taking the appropriate security decision requires access to the policies. However, if such access is allowed in an untrusted environment then confidential information might be leaked by the policies. Current solutions are based on cryptographic operations that embed security policies with the security mechanism. Therefore, the enforcement of such policies is performed by allowing the authorised parties to access the appropriate keys. We believe that such solutions are far too rigid because they strictly intertwine authorisation policies with the enforcing mechanism.

In this paper, we want to address the issue of enforcing security policies in an untrusted environment while protecting the policy confidentiality. Our solution ESPOON is aiming at providing a clear separation between security policies and the enforcement mechanism. However, the enforcement mechanism should learn as less as possible about both the policies and the requester attributes.

\end{abstract}


\IEEEpeerreviewmaketitle

\section{Introduction}

The cost saving associated with a general improvement in the quality of services and operations provided makes outsourcing of the IT infrastructure a business model adopted by many companies. Even sectors such as healthcare initially reluctant to this model are now slowly adopting it \cite{Ondo06}. Outsourcing typically relies on third parties to provide and maintain a very reliable IT infrastructure. However, the data stored on the outsourced servers are within easy reach of the infrastructure provider that could reuse the data for unintended and/or malicious purposes.

Several technical approaches have been proposed to guarantee the confidentiality of the data in an outsourced environment. For instance, solutions as those described in \cite{Dong08, Kamara10} allow a protected storage of data while maintaining basic search capabilities to be performed on the server side. However, such solutions do not support access policies to regulate the access of a user (or a group of users) to a particular subset of the stored data.

\subsection{Motivation}

Solutions for providing access control mechanisms in outsourced environments have mainly focused on encryption techniques that couple access policies with set of keys, such as the one described in \cite{Vimercati08}. Only users possessing a key (or a set of hierarchy-derivable keys) are authorised to access the data. The main drawback of these solutions is that security policies are tightly coupled with the security mechanism incurring in high processing cost for performing any administrative change for both the users and the policies representing the access rights.

A policy-based solution such the one described for the Ponder language in \cite{russello07} results more flexible and easy to manage because it clearly separates the security policies from the enforcement mechanism. However, policy-based access control mechanisms were not designed to operate in outsourced environments. Such solution can work only when they are deployed and operated within a trusted domain (i.e., the computational environment managed by the organisation owning the data). If these mechanisms are outsourced in an untrusted environment, the access policies that are to be enforced on the server may leak information on the data they are protecting. As an example, let us consider a scenario where a hospital has outsourced its healthcare data management services to a third party service provider. We assume that the service provider is honest-but-curious, similarly to the existing literature on data outsourcing. That is, it is honest to perform the required operations as described in the protocol but curious to know the data contents. In other words, the service provider does not preserve the data confidentiality. A patient's medical record should be associated with an access policy to prevent that any hospital employees is allowed to see the patient's data. The data is stored with an access policy on the outsourced environment. As an example of such an access policy, let us consider the following access policy: \emph{only a Cardiologist may access the data}. From this policy, it is possible to infer important information about the user's medical data (even if the actual medical record is encrypted). This policy reveals that a patient could have heart problems. A curious service provider may sell this information to banks that could deny the patient a loan given her health condition.

\subsection{Research Contributions}

In this paper, we present a policy-based access control mechanism for outsourced environments where we support full confidentiality of the access policies. We named our solution \textbf{ESPOON} (Encrypted Security Policies for OutsOursed eNvironments). One of the main advantages of ESPOON is that we maintain the clear separation between the security policies and the actual enforcing mechanism without loss of confidentiality. This can be guaranteed under the assumption that the service provider is honest-but-curious. Our approach allows to implement the access control mechanism as an outsourced service with all the benefits associated with this business model without compromising the confidentiality of the policies. Summarising, the research contributions of our approach are threefold. First of all, the service provider does not learn anything about policies and the requester's attributes during the policy evaluation process. Second, ESPOON is capable of handling complex policies involving non-monotonic boolean expressions and range queries. Third, the system entities do not share any encryption keys and even if a user is deleted or revoked, the system is still able to perform its operations without requiring re-encryption of the policies. As a proof-of-concept, we have implemented a prototype architecture of our access control mechanism and analysed its performance to quantify the incurred overhead.

\subsection{Organisation}
The rest of this paper is organised as follows: Section \ref{sec:related_work} reviews the related work. Section \ref{sec:proposed_solution} presents the proposed architecture of ESPOON. Section \ref{sec:solution_details} focuses on implementation details involved in ESPOON. Section \ref{sec:discussion} provides the discussion about the security and privacy aspects of ESPOON. Section \ref{sec:performance_evaluation} analyses the performance of ESPOON. Finally, Section \ref{sec:conclusion_future_work} concludes this paper and gives directions for the future work.

\section{Related Work}
\label{sec:related_work}
Work on outsourcing data storage to a third party has been focusing on protecting the data confidentiality within the outsourced environment. Several techniques have been proposed allowing authorised users to perform efficient queries on the encrypted data while not revealing information on the data and the query \cite{Song00, Boneh04, Golle04, Curtmola06, Hwang07, Boneh07, Wang08, Baek08, Dong08, Rhee10, Shao10}. However, these techniques do not support the case of users having different access rights over the protected data. Their assumption is that once a user is authorised to perform search operations, there are no restrictions on the queries that can be performed and the data that can be accessed.

The idea of using an access control mechanism in an outsourced environment was initially explored in \cite{Vimercati07, Vimercati07-2}. In this approach, the authors provide a selective encryption strategy as a means for access control enforcement. The idea is to have a selective encryption technique where each user has a different key capable of decrypting only the resources a user is authorised to access. In their scheme, a public token catalogue expresses key derivation relationships. However, the public catalogue contains tokens in the clear that express the key derivation structure. The tokens could leak information on the security policies and on the protected data. To circumvent the issue of information leakage, in \cite{Vimercati08} the authors provide an encryption layer to protect the public token catalogue. This requires each user to obtain the key for accessing a resource by traversing the key derivation structure. The key derivation structure is a graph built (using access key hierarchies \cite{Atallah09}) from a classical access matrix. There are several issues related to this scheme. First, the algorithm of building key derivation structure is very time consuming. Any administrative actions to update access rights require the users to obtain new access keys derived from the rebuilt key derivation structure and it consequently requires data re-encryption with new access keys. Therefore, the scheme is not very scalable and may be suitable for a static environment where users and resources do not change very often. Last but not least, the scheme does not support complex policies where contextual information may be used for granting access rights. For instance, only specific time and location information associated with an access request may be legitimate to grant access to a user.

Another possible approach for implementing an access control mechanism is protecting the data with an encryption scheme where the keys can be generated from the user's credentials (expressing attributes associated with that user). Although these approaches were not devised particularly for outsourced environments, it is still possible to use them as access control mechanisms in outsourced settings. For instance, a recent work by Narayan et al. \cite{Narayan10} employ the variant of Attribute Based Encryption (ABE) proposed in \cite{Bethencourt07} (that is the Ciphertext Policy ABE, or CP-ABE in short) to construct an outsourced healthcare system where patients can securely store their Electronic Health Record (EHR). In their solution, each EHR is associated with a secure search index to provide search capabilities while guaranteeing no leakage of information. However, one of the problems associated with CP-ABE is that the access structure, representing the security policy associated with the encrypted data, is not protected. Therefore, a curious storage provider might get information on the data by accessing the attributes expressed in the CP-ABE policies. The problem of having the access structure expressed in cleartext affects in general all the ABE constructions \cite{Sahai05,Goyal06,Ostrovsky07,Bethencourt07}. Therefore, this mechanism is not suited for protecting the confidentiality of the access policies in an outsourced environment.

Related to the issue of the confidentiality of the access structure, the hidden credentials scheme presented in \cite{Holt03} allows to decrypt ciphertexts while the parties involved never reveal their own policies and credentials to each other. Data can be encrypted using an access policy containing monotonic boolean expressions which must be satisfied by the receiver to get access to the data. A passive adversary may deduce the policy structure, i.e., the operators (AND, OR, m-of-n threshold encryption) used in the policy but she does not learn what credentials are required to fulfil the access policy unless she possesses them. Bradshaw et al. \cite{Bradshaw04} extend the original hidden credentials scheme to limit the partial disclosure of the policy structure and speed up the decryption operations. However, in this scheme is not easy to support non-monotonic boolean expressions and range queries in the access policy. Last, hidden credentials schemes assume that the involved parties have to be online all the time to run the protocol.

\begin{figure} [ht]
\centering
% left bottom right top
\includegraphics[trim=100mm 75mm 90mm 65mm,clip,width=.35\textwidth]{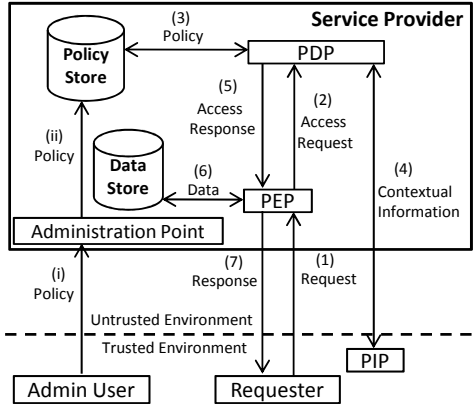}
\caption{The ESPOON architecture for policy enforcement in outsourced environments}
\label{fig:abstract_picture}
\end{figure}

\section{The ESPOON Approach}
\label{sec:proposed_solution}

ESPOON aims at providing a policy-based access control mechanism that can be deployed in an outsourced environment. Figure \ref{fig:abstract_picture} illustrates the proposed architecture that has similar components as the widely accepted architecture for policy-based management proposed by IETF \cite{rfc2753}. In ESPOON, the \textbf{Admin User} deploys (i) the access policies to the \textbf{Administration Point} that stores (ii) the policies in the \textbf{Policy Store}. Whenever a \textbf{Requester}, say a doctor, needs to access the data, a request is sent to the \textbf{Policy Enforcement Point} (PEP) (1). This request includes the Requester's identifier (subject), the requested data (target) and the action to be performed. The PEP (2) forwards the access request to the \textbf{Policy Decision Point} (PDP). The PDP (3) obtains the policies matching against the access request from the \textbf{Policy Store} and (4) retrieves the contextual information from the \textbf{Policy Information Point} (PIP). The contextual information may include the environmental and Requester's attributes under which an access can be considered valid. For instance, a doctor should only access the data during the office hours. For simplicity, we assume that the PIP collects all required attributes including the Requester's attributes and sends all of them together in one go. Moreover, we assume that the PIP is deployed in the trusted environment. However if attributes forgery is an issue, then the PIP can request a trusted authority to sign the attributes before sending them to the PDP. The PDP evaluates the policies against the attributes provided by the PIP checking if the contextual information satisfies any policy conditions and sends to the PEP the access response (5). In case of \emph{permit}, the PEP forwards the access action to the Data Store (6). Otherwise, in case of \emph{deny}, the requested action is not forwarded. Optionally, a response can be sent to the Requester (7) with either \emph{success} or \emph{failure}.

The main difference with the standard proposed by IETF is that the ESPOON architecture for policy-based access control is outsourced in an untrusted environment (see Figure \ref{fig:abstract_picture}). The trusted environment comprises only a minimal IT infrastructure that is the applications used by the Admin Users and Requesters, together with the PIP. This reduces the cost of maintaining an IT infrastructure. Having the reference architecture in the cloud increases its availability and provides a better load balancing compared to a centralised approach. Additionally, ESPOON guarantees that the confidentiality of the policies is protected while their evaluation is executed on the outsourced environment. This allows a more efficient evaluation of the policies. For instance, a naive solution would see the encrypted policies stored in the cloud and the PDP deployed in the trusted environment. At each evaluation, the encrypted policies would be sent to the PDP that decrypts the policies for a cleartext evaluation. After that, the policies need to be encrypted and send back to the cloud. The \textbf{Service Provider} where the architecture is outsourced is honest-but-curious. This means that the provider allows the ESPOON components to follow the specified protocols, but it may be curious to find out information about the data and the policies regulating the accesses to the data. As for the data, we assume that the confidentiality data is protected by one of the several techniques available for outsourced environments (see \cite{Dong08, Rhee10, Shao10}). However, to the best of our knowledge no solution exists that addresses the problem of guaranteeing policy confidentiality while allowing an efficient evaluation mechanism that is clearly separated from the security policies. Most of the techniques discussed in the related work section require the security mechanism to be tightly coupled with the policies. In the following section, we can show that it is possible to maintain a generic PDP separated from the security policies and able to take access decisions based on the evaluation of encrypted policies. In this way, the policy confidentiality can be guaranteed against a curious provider and the functionality of the access control mechanism is not restricted.

\subsection{System Model}
Before presenting the details of the scheme used in ESPOON, it is necessary to discuss the system model. In this section, we identify the following system entities.
%In this section, we first identify the entities involved and the assumptions underlying the system design. Then we identify the potential adversaries and the possible attacks. The main types of system entities are:

\begin{itemize}
\item \textbf{Admin User:} This type of user is responsible for the administration of the policies stored in the outsourced environment. An Admin User can deploy new policies on the outsourced environment or update/delete the policies already deployed.
\item \textbf{Requester:} A Requester is a user that requests an access (e.g., read, write, search, etc.) over the data residing on the outsourced environment. Before the access is permitted, the policies deployed on the outsourced environment are evaluated.
% TODO: should we put PIP here? If yes, why not PDP, PEP, Administration Point?
\item \textbf{Service Provider (SP):} The SP is responsible for managing the outsourced computation environment, where the ESPOON components are deployed and to store the data, and the access policies. It is assumed the SP is honest-but-curious, i.e, it allows the components to follow the protocol to perform the required actions but curious to deduce information about the exchanged and stored policies.
\item \textbf{Trusted Key Management Authority (KMA):} The KMA is fully trusted and responsible for generating and revoking the keys. For each type of authorised users (both the Admin User and Requester), the KMA generates a key pair and securely transmits one part of the generated key pair to the user and the other to the SP. The KMA is deployed on the trusted environment. Although requiring a Trusted KMA seems at odds with the needs of outsourced the IT infrastructure, we argue that the KMA requires less resources and less management effort. Securing the KMA is much easier since a very limited amount of data needs to be protected and the KMA can be kept offline most of time.
\end{itemize}

It should be clarified that in our settings an Admin User is not interested in protecting the confidentiality of access policies from other Admin Users and Requesters. Here, the main goal is to protect the confidentiality of the access policies from the SP.

\begin{figure} [htp]
\centering
% left bottom right top
\includegraphics[trim=85mm 70mm 85mm 50mm,clip,width=.3445\textwidth]{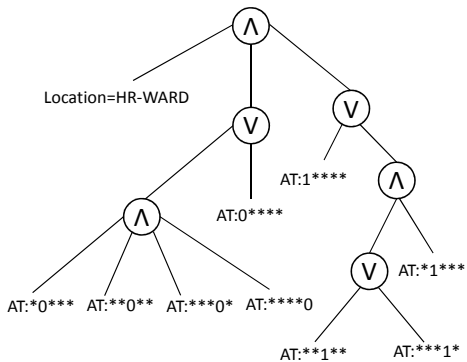}
\caption{Example of access policy condition tree illustrating $Location=HR \mhyphen WARD$ and $AT>9\#5$ and $AT<17\#5$}
\label{fig:cc}
\end{figure}

\subsection{Policy Representation}
In this section, we provide an informal description of the policy representation used in our approach. For this paper, we deal with only positive authorisation policies. This means that, as default no actions are allowed unless at least one authorisation policy can be applicable to the request.

In our approach, an authorisation policy is represented as follows:

\begin{small}
 $IF$ condition $THEN$ $CAN$ $\langle S, A, T \rangle$
\end{small}

The meaning is the following: if the condition is true then the subject $S$ can execute the action $A$ on the target $T$. At the time when a request is made, the information about the subject, the action that is requested and the target resource is collected by the Requester. The PIP collects several attributes representing the context in which the request is being executed and sends them to the PDP.

To represent the condition in a policy, we use the tree structure described in \cite{Bethencourt07} for CP-ABE policies. This access tree structure allows to express conjunctions and disjunctions of equalities and inequalities. Internal nodes of the tree are AND, OR or threshold gates (e.g., 2 of 3) and leaf nodes are the values of the condition predicates. To support comparisons between numerical values a representation of ``bag of bits'' can be used. For instance, let us take a condition stating that the Requester location should be the HR ward and that the time of the access should be between 9:00 and 17:00. The tree representing such condition can be built using AND and OR gates as shown in Figure \ref{fig:cc}. When the request is made, the location of the Requester is the HR ward and the access time is 10:00. To make things simpler, let us concentrate only on the numerical value representing the hour. The following attributes will be sent to the PDP by the PIP:

\begin{small}
(``Location=HR-WARD'', ``AT : 0****'', ``AT: *1***'',``AT: **0**'', ``AT: ***1*'', ``AT: ****0'')
\end{small}.
The first attribute represents the location while the other attributes are used for representing the time of the access, i.e., the value 10 in a 5-bit representation. Basically, the attributes are matched against the leaf nodes of the tree according to the AND and OR gates.

In this policy representation, the $\langle S, A, T \rangle$ tuple and the leaf nodes in the condition tree are in clear text. Therefore, such information is easily accessible on the outsourced environment and may leak information about the data that the policies protect. In the following, we show how such representation can be protected while allowing the PDP to evaluate the policies against the request.

%In the following, we show how we protect the Requester attributes and the leaf nodes in the access policy condition tree while allowing the PDP to evaluate the policies. However, we do not protect the access structure i.e., one may deduce the information about the operators, such as AND and OR, used in the access policy condition.

\section{Solution Details}
\label{sec:solution_details}

The main idea of our approach is to use an encryption scheme for protecting the confidentiality of the policies while allowing the PDP to perform the correct evaluation of the policies. We noticed that the operation performed by the PDP for evaluating policies is similar to the search operation executed in a database. In particular, in our case the condition of a policy is the query; and the data that is matched against the query is represented by the attributes that the Requester sends in the request.

For this reason, as a starting point we have taken a multi-user searchable data (SDE) scheme as the one proposed in \cite{Dong08} that allows an untrusted server to perform searches over encrypted data without revealing to the server information on both the data and the elements used in the request. The advantage of this method is that it allows multi-user access without the need for a shared key between users. Each user in the system has a unique set of keys. The data encrypted by one user can be decrypted by any other authorised user. However, the SDE implementation in \cite{Dong08} is only able to perform keyword comparison based on equalities. One of the major extensions of our implementation is that we are able to support the evaluation of conditions with complex boolean expressions such as non-conjunctive and range queries in the multi-user settings.

In general, in a policy-based approach it is possible to distinguish two main phases in the policy life cycle: the first phase is the \textbf{policy deployment} in to the Policy Store; and the second phase is the \textbf{policy evaluation} when a request is made. In the following, we provide the details of the algorithms based on the SDE scheme proposed in \cite{Dong08} used in each phase.

\subsection{Initialisation Phase}
Before the policy deployment and policy evaluation phases, the SDE scheme needs to be initialised. This is done for generating the required keying material. The following two functions need to be called:

\begin{itemize}

\item The initialisation algorithm $Init(1^k)$ is run by the Trusted KMA. It takes as input the security parameter $1^k$ and outputs the public parameters $Params$ and the master secret key set $MSK$.

\item The user key sets generation algorithm $KeyGen(MSK, i)$ is run by the Trusted KMA. It takes as input the master secret key set $MSK$ and the user (Admin User or Requester) identity $i$ and generates two key sets $K_{u_i}$ and $K_{s_i}$. The Trusted KMA sends key sets $K_{u_i}$ and $K_{s_i}$ to the user $i$ and the Key Store, respectively. Only the Administration Point, PDP and PEP are authorised to access the Key Store.
\end{itemize}

\subsection{Policy Deployment Phase}
The policy deployment phase is executed when a new set of policies needs to be deployed on the Policy Store (or an existing version of policies needs to be updated). This phase is executed by the Admin User who edits the policies in a trusted environment. Before the policies leave the trusted environment, they need to be encrypted. Our policy representation consists of two parts: one for representing the condition and the other for the $\langle S, A, T \rangle$ tuple. Each part is encrypted by the following functions:

\begin{itemize}
\item The access policy condition encryption algorithm $PD \mhyphen Condition \mhyphen Enc(Condition, K_{u_i})$ is run by the Admin User $i$. It takes as input $Condition$ and $K_{u_i}$ and outputs the ciphertext $c^*_i (Condition)$.

\item The access policy $\langle S, A, T \rangle$ tuple encryption algorithm $PD \mhyphen SAT \mhyphen Enc(\langle S, A, T \rangle, K_{u_i})$ is run by the Admin User $i$. It takes as input the $\langle S, A, T \rangle$ tuple and $K_{u_i}$ and outputs the ciphertext $c^*_i (\langle S, A, T \rangle)$.
\end{itemize}

When the encrypted policy is sent to the outsourced environment, then another encryption round is performed. This is done using the following functions:

\begin{itemize}

\item The access policy condition re-encryption algorithm $PD \mhyphen Condition \mhyphen Re \mhyphen Enc(i, c^*_i (Condition), K_{s_i})$ is run by the Administration Point. It takes as input $c^*_i (Condition)$ and the key $K_{s_i}$ corresponding to the Admin User $i$ and outputs the re-encrypted ciphertext $c(Condition)$.

\item The access policy $\langle S, A, T \rangle$ tuple re-encryption algorithm $PD \mhyphen SAT \mhyphen Re \mhyphen Enc(c^*_i (\langle S, A, T \rangle), K_{u_i})$ is run by the Admin User $i$. It takes as input the encrypted $\langle S, A, T \rangle$ tuple $c^*_i (\langle S, A, T \rangle)$ and the key $K_{s_i}$ corresponding to the Admin User $i$ and outputs the re-encrypted ciphertext $c(\langle S, A, T \rangle) = \{ c(S), c(A), c(T) \}$.

\end{itemize}

The access policy can be now stored in the Policy Store. The stored policies do not reveal any information about the data because these are stored as encrypted.

\subsection{Policy Evaluation Phase}
The policy evaluation phase is executed when a Requester makes a request to access the data. Before the access permission, the PDP evaluates the matching policies in the Policy Store on the SP. The request contains the $\langle S, A, T \rangle$ tuple. This information is encrypted by the following function before it leaves the trusted environment:

\begin{itemize}

\item The access policy $\langle S, A, T \rangle$ tuple encryption algorithm $PE \mhyphen SAT \mhyphen Enc(\langle S, A, T \rangle, K_{u_j})$ is run by the Requester $j$. It takes as input the $\langle S, A, T \rangle$ tuple and $K_{u_j}$ and outputs the encrypted $\langle S, A, T \rangle$ tuple $T^*_j (\langle S, A, T \rangle)$.

\end{itemize}

% As a last step of the policy evaluation phase at the Requester side, the encrypted attributes and the $\langle S, A, T \rangle$ tuple from the Requester are sent to outsourced environment. The policy evaluation phase at the SP side starts with searching all the policies in the Policy Store matching against the Requester $\langle S, A, T \rangle$ tuple. This is accomplished by the following function:

The Requester sends the encrypted $\langle S, A, T \rangle$ tuple to the SP. The policy evaluation phase at the SP side starts with searching all the policies in the Policy Store matching against the Requester $\langle S, A, T \rangle$ tuple. This is accomplished by the following function:

\begin{itemize}

% revise alignment if the following paragraph is updated
\item The $\langle S, A, T \rangle$ tuple search algorithm $PE \mhyphen SAT \mhyphen Search(j, T^*_j (\langle S, A, T \rangle), K_{s_j},$ $c(\langle S_i, A_i, T_i \rangle)_{1 \leq i \leq n})$ is run by the PDP. It takes as input the encrypted $\langle S, A, T \rangle$ tuple $T^*_j (\langle S, A, T \rangle)$ from the Requester $j$, the key $K_{s_j}$ corresponding to the Requester $j$, and all $n$ stored policies in the Policy Store $c(\langle S_i, A_i, T_i \rangle)_{1 \leq i \leq n}$ and returns the matching tuples in the Policy Store.

\end{itemize}

If any match is found in the Policy Store then the PDP needs to match the contextual information against the access policy condition corresponding to the matched tuple. The PDP fetches the contextual information including Requester and environmental attributes from the PIP. The PIP encrypts the contextual information by the following function before it leaves the trusted environment:

\begin{itemize}

\item The attributes encryption algorithm $PE \mhyphen Attributes \mhyphen Enc(\gamma, K_{u_j})$ is run by the PIP $j$. It takes as input the Requester and environmental attributes $\gamma$ and $K_{u_j}$ and outputs the encrypted attributes $T_j (\gamma)$.

\end{itemize}

After receiving the contextual information from the PIP, the PDP matches the PIP attributes against the access policy condition. The PDP calls the following function to evaluate the access policy condition:

\begin{itemize}

\item The access policy condition evaluation algorithm $PE \mhyphen Condition \mhyphen Evaluation(j, T_j (\gamma), K_{s_j}, c(Condition))$ is run by the PDP. It takes as input a list of encrypted attributes $T_j (\gamma)$ and the key $K_{s_j}$ both corresponding to the PIP $j$ and encrypted access policy condition tree $c(Policy)$ and outputs either $Permit$ or $Deny$.
\end{itemize}

\subsection{Revocation}
The proposed solution allows revocation of a user (Admin User or Requester). For this purpose, the Administration Point calls the following function:
\begin{itemize}
\item The user (Admin User or Requester) revocation algorithm $Revoke(i)$ is run by the Administration Point. Given the user $i$, the Administration Point removes the corresponding key $K_{s_i}$ from the Key Store.
\end{itemize}

\subsection{Concrete Construction}
This section provides the cryptographic details of each function in the phases described above:

\begin{figure} [ht]
\centering
% left bottom right top
\includegraphics[trim=95mm 75mm 95mm 65mm,clip,width=.3\textwidth]{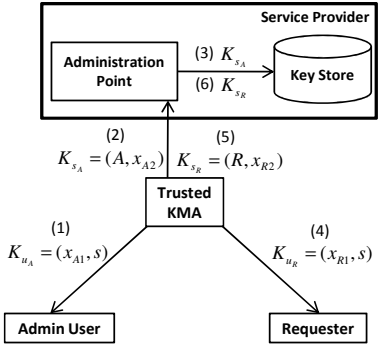}
\caption{Key distribution}
\label{fig:general_picture}
\end{figure}

\begin{itemize}

\item $Init(1^k):$ The Trusted KMA takes as input the security parameter $1^k$ and outputs two prime numbers $p$, $q$ such that $q$ divides $p - 1$, a cyclic group $\mathbb{G}$ with a generator $g$ such that $\mathbb{G}$ is the unique order $q$ subgroup of $\mathbb{Z}^*_P$. It chooses $x \xleftarrow{R} \mathbb{Z}^*_q$ and compute $h = g^x$. Next, it chooses a collision-resistant hash function $H$, a pseudorandom function $f$ and a random key $s$ for $f$. Finally it publicises the public parameters $Params = (\mathbb{G}, g, q, h, H, f)$ and keeps securely the master secret key $MSK = (x, s)$.

\item $KeyGen(MSK, i):$ For each user (Admin User or Requester) $i$, the Trusted KMA chooses $x_{i1} \xleftarrow{R} \mathbb{Z}^*_q$ and computes $x_{i2} = x - x_{i1}$. It securely transmits $K_{u_i} = (x_{i1}, s)$ to the user $i$ and $K_{s_i} = (i, x_{i2})$ to the Administration Point which inserts $K_{s_i}$ in the Key Store, i.e., $K_S = K_S \cup K_{s_i}$\footnote{The Key Store is initialised as $K_S = \Phi$.}. Figure \ref{fig:general_picture} illustrates the key distribution process where an Admin User say $A$ and a Requester say $R$ are receiving their keys (1) $K_{u_A}$ and (4) $K_{u_R}$, respectively. The corresponding SP side keys (2) $K_{s_A}$ and (5) $K_{s_R}$ are sent to the Administration Point that stores both (3) $K_{s_A}$ and (6) $K_{s_R}$ in the Key Store.

\begin{figure} [ht]
\centering
% left bottom right top
\includegraphics[trim=100mm 55mm 95mm 50mm,clip,width=.3\textwidth]{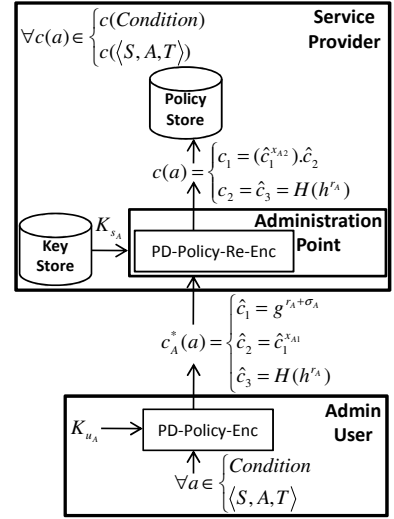}
\caption{ESPOON: Policy deployment phase}
\label{fig:patient}
\end{figure}

% TODO: optional: add refernce to tree here: threshold gate, and gate, or gate, -ve values
% TODO: for journal version?
\item $PD \mhyphen Condition \mhyphen Enc(Condition, K_{u_i}):$ The Admin User defines the access policy condition as a tree $Condition$ denoting a set of string and numerical comparisons represented by $n$ leaf nodes. Each non-leaf node $a'$ in $Condition$ represents a threshold gate with the threshold value $k_{a'}$ denoting the number of its child subtrees that must be satisfied where $a'$ has total $c_{a'}$ child subtrees, i.e., $1 \leq k_{a'} \leq c_{a'}$. If $k_{a'} = 1$, the threshold gate is an OR and if $k_{a'} = c_{a'}$, the threshold gate is an AND. Each leaf node $a$ represents either a string comparison or subpart of a numerical comparison (because one numerical comparison of size $s$ bits is represented by $s$ leaf nodes at the most) with a threshold value $k_{a} = 1$. For every leaf node $a \in Condition$, the Admin User chooses $r_{a} \xleftarrow{R} \mathbb{Z}^*_q$ and computes $c^*_i (a) = (\hat{c}_1, \hat{c}_2, \hat{c}_3)$ where $\hat{c}_1 = g^{r_a+{\sigma}_a}, {\sigma}_a = f_s (a), \hat{c}_2 = \hat{c}_1^{x_{i1}}, \hat{c}_3 = H(h^{r_a})$. The Admin User transmits to the Administration Point the encrypted access policy condition tree $c^*_i (Condition) = \{c^*_i (a_1), c^*_i (a_2), \ldots, c^*_i (a_n)\}$ as shown in Figure \ref{fig:patient}.

\item $PD \mhyphen SAT \mhyphen Enc(\langle S, A, T \rangle, K_{u_i}):$ The Admin User defines and encrypts each item in the $\langle S, A, T \rangle$ tuple. For each item $a \in \langle S, A, T \rangle$, the Admin User chooses $r_{a} \xleftarrow{R} \mathbb{Z}^*_q$ and computes $c^*_i (a) = (\hat{c}_1, \hat{c}_2, \hat{c}_3)$ where $\hat{c}_1 = g^{r_a+{\sigma}_a}, {\sigma}_a = f_s (a), \hat{c}_2 = \hat{c}_1^{x_{i1}}, \hat{c}_3 = H(h^{r_a})$. The Admin User transmits to the Administration Point the encrypted $\langle S, A, T \rangle$ tuple $c^*_i (\langle S, A, T \rangle) = \{c^*_i (S), c^*_i (A), c^*_i (T)\}$ as shown in Figure \ref{fig:patient}.

\item $PD \mhyphen Condition \mhyphen Re \mhyphen Enc(i, c^*_i (Condition), K_{s_i}):$ The Administration Point retrieves the key $K_{s_i}$ corresponding to the Admin User $i$. Each encrypted leaf node $c^*_i (a)$ in $c^*_i (Condition)$ is re-encrypted to $c(a) = (c_1, c_2)$, where $c_1 = (\hat{c}_1)^{x_{i2}}.\hat{c}_2 = \hat{c}_1^{x_{i1}+x_{i2}} = (g^{r_a+{\sigma}_a})^x = h^{r_a+{\sigma}_a}$ and $c_2 = \hat{c}_3 = H(h^{r_a})$. The Administration Point stores the re-encrypted access policy condition $c(Condition) = \{c(a_1), c(a_2), \ldots, c(a_n)\}$ in the Policy Store as shown in Figure \ref{fig:patient}.

\item $PD \mhyphen SAT \mhyphen Re \mhyphen Enc(c^*_i (\langle S, A, T \rangle), K_{u_i}):$ The Administration Point retrieves the key $K_{s_i}$ corresponding to the Admin User $i$. Each encrypted item $c^*_i (a)$ in $c^*_i (\langle S, A, T \rangle)$ is re-encrypted to $c(a) = (c_1, c_2)$, where $c_1 = (\hat{c}_1)^{x_{i2}}.\hat{c}_2 = \hat{c}_1^{x_{i1}+x_{i2}} = (g^{r_a+{\sigma}_a})^x = h^{r_a+{\sigma}_a}$ and $c_2 = \hat{c}_3 = H(h^{r_a})$. The Administration Point stores the re-encrypted $\langle S, A, T \rangle$ tuple $c(\langle S, A, T \rangle) = \{ c(S), c(A), c(T) \}$ in the Policy Store as shown in Figure \ref{fig:patient}.

\begin{figure} [ht]
\centering
% left bottom right top
\includegraphics[trim=80mm 80mm 85mm 50mm,clip,width=.4\textwidth]{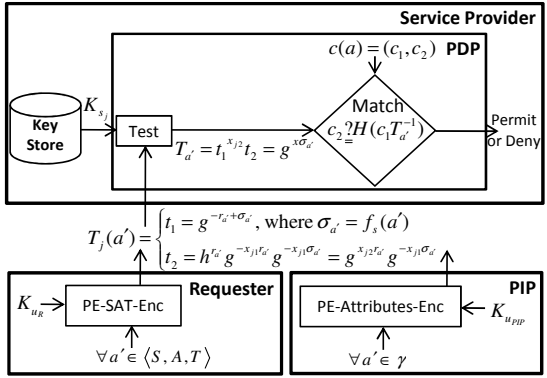}
\caption{ESPOON: Policy evaluation phase}
\label{fig:doctor}
\end{figure}

\item $PE \mhyphen SAT \mhyphen Enc(\langle S, A, T \rangle, K_{u_j}):$ To encrypt each item $a'$ in the $\langle S, A, T \rangle$ tuple, the Requester $j$ chooses $r_{a'} \xleftarrow{R} \mathbb{Z}^*_q$ and computes trapdoor $T_j (a') = (t_1, t_2)$ where $t_1 = g^{-r_{a'}} g^{{\sigma}_{a'}}$ and $t_2 = h^{r_{a'}} g^{-x_{j1}r_{a'}} g^{x_{j1}{\sigma}_{a'}} = g^{x_{j2}r_{a'}} g^{x_{j1}{\sigma}_{a'}}$, where ${\sigma}_{a'} = f_s (a')$. She encrypts the $\langle S, A, T \rangle$ tuple as $T_j (\langle S, A, T \rangle) = \{ T^*_j (S), T^*_j (A), T^*_j (T) \}$ as shown in Figure \ref{fig:doctor} and sends it to the PDP.

% if aligned then update it
\item $PE \mhyphen SAT \mhyphen Search(j, T^*_j (\langle S, A, T \rangle), K_{s_j}, c(\langle S_i, A_i,$ $T_i \rangle)_{1 \leq i \leq n}):$ The PDP receives the access request containing encrypted $\langle S, A, T \rangle$ tuple $T^*_j (\langle S, A, T \rangle)$. Next, it retrieves the key $K_{s_j}$ corresponding to the Requester $j$. For each $c(\langle S, A, T \rangle) \in c(\langle S_i, A_i, T_i \rangle)_{1 \leq i \leq n}$, it matches $c(\langle S, A, T \rangle)$ with $T^*_j (\langle S, A, T \rangle)$. To perform this matching, it has to check if each trapdoor $T^*_j (a') \in T^*_j (\langle S, A, T \rangle)$ is matching against the corresponding item $c(a) \in c(\langle S, A, T \rangle)$. The match is accomplished as follows. The PDP computes $T = t_1^{x_{j2}} . t_2 = g^{x{\sigma}_{a'}}$ as shown in Figure \ref{fig:doctor}. Next, it tests if $c_2 \stackrel{?}{=} H(c_1 . T^{-1})$. If so, then trapdoor $T^*_j (a')$ is considered matched against $c(a)$. If all trapdoors of $T^*_j (\langle S, A, T \rangle)$ matching against the corresponding items of $c(\langle S, A, T \rangle)$, the tuple will be considered matched and this function returns all matching $\langle S, A, T \rangle$ tuples in the Policy Store.

\item $PE \mhyphen Attributes \mhyphen Enc(\gamma, K_{u_j}):$ To encrypt each attribute $a'$ where $a' \in \gamma$, the PIP\footnote{The key generation for the PIP is similar to that of a Requester.} $j$ chooses $r_{a'} \xleftarrow{R} \mathbb{Z}^*_q$ and computes trapdoor $T_j (a') = (t_1, t_2)$ where $t_1 = g^{-r_{a'}} g^{{\sigma}_{a'}}$ and $t_2 = h^{r_{a'}} g^{-x_{j1}r_{a'}} g^{x_{j1}{\sigma}_{a'}} = g^{x_{j2}r_{a'}} g^{x_{j1}{\sigma}_{a'}}$, where ${\sigma}_{a'} = f_s (a')$. The PIP encrypts all attributes $\gamma$ as $T_j (\gamma) = \{ T_j(a'_1), T_j(a'_2), \ldots, T_j(a'_m) \}$ as shown in Figure \ref{fig:doctor} and sends it to the PDP.

\item $PE \mhyphen Condition \mhyphen Evaluation(j, T^*_j (\gamma), K_{s_j}, c(Condition)):$ The PDP receives the contextual information containing trapdoors $T_j (\gamma)$ from the PIP $j$. Next, it retrieves the key $K_{s_j}$ corresponding to the PIP $j$ and the encrypted access policy condition $c(Condition)$. The PDP performs recursive algorithm starting from the root node of the access policy condition tree $Condition$. For each non-leaf node, it checks if the number of children that are satisfied is greater than or equal to the threshold value of the node. If so, the node is marked as satisfied. For each encrypted leaf node $c(a)$ in $c(Condition)$, there may exist a corresponding PIP trapdoor $T_j(a')$ where $T_j(a') \in T_j(\gamma)$. For this purpose, it computes $T = t_1^{x_{i2}} . t_2 = g^{x{\sigma}_{a'}}$ as shown in Figure \ref{fig:doctor}. For each leaf node $c(a) = (c_1, c_2)$, it tests if $c_2 \stackrel{?}{=} H(c_1 . T^{-1})$. If so, the leaf node is marked as satisfied. Its response to the PEP is $Permit$ if the root node of the access policy condition tree $Condition$ is marked as satisfied and $Deny$ otherwise.

\item $Revoke(i):$ The Administration Point removes the key $K_{s_i}$ corresponding to the user (Admin User or Requester) $i$ from the Key Store as $K_S = K_S \setminus K_{s_i}$. Therefore, both the PEP and the Administration Point check the revocation of a user before invoking any action.

\end{itemize}

% TODO: Mention about this section in the organisation of this paper
\section{Discussion}
\label{sec:discussion}
%In this section, we discuss various extensions to ESPOON?
This section provides the discussion about the security and privacy aspects of ESPOON.

\subsection{On the Impossibility of Cryptography Alone for Privacy-Preserving Cloud Computing}
Dijk and Juels argue in \cite{VanDijk2010} that cryptography alone is not sufficient for preserving the privacy in the cloud environment. They prove that in multi-client settings it is impossible to control how information is released to clients with different access rights. Basically, in their threat model clients do not mutually trust each other. In our settings, users are mutually trusted: our main contribution is to protect the confidentiality of the access policies (and therefore of the data) from the SP.

\subsection{Revealing Policy Structure}
The access policy structure reveals information about the operators, such as AND and OR, and the number of operands used in the access policy condition. To overcome this problem, dummy attributes may be inserted in the tree structure of the access policy. Similarly, the PIP can send dummy attributes to the PDP at the time of policy evaluation to obfuscate the number of attributes required in a request.

\iffalse
\subsection{Collusion Attack}
In ESPOON, a single compromised user $i$ (either Admin User or Requester) may recover the master secret key $x$ by colluding with the SP. To withstand the collusion attack, secret user key $K_{U_i}$ can be split in to two parts. One part is given to the user $i$ while the other part is managed by the application gateway to access the SP. For instance, in case where a hospital has outsourced its IT services to a SP, the gateway is managed by the hospital to access the SP. An application gateway manages a User Key Store to store part of the user keys. The Trusted KMA is responsible for user key splitting and sending the corresponding shares directly to the user $i$ and the application gateway.
\fi

%\subsection{Security Proofs}
%Policy indistinguishability? \\ % \cite{Bradshaw04}
%Attribute indistinguishability? % \cite{Bradshaw04}
% TODO: for journal version?

% Theorems, lemmas, definitions, security assumption, security analysis, Time complexity or efficiency

%% Null policy means no policy. always satisfied.
%% TODO: decide which one. SHA-1?
%%R_d = D_a \cup Environment_a \cup Context_a

%%\subsection{Complexity}
%%For each entry in encrypted policy, say $P_p$ ($F_s$ in case of pub/sub), there must be an attribute (i.e. trapdoor). Let say there are $n_a$ number of trapdoors and $n_b$ number of entries in the access tree, the complexity whether a doctor satisfies the policy, is $O(n_a n_b)$.
%%\begin{eqnarray}
%%|TD_d (a), a \in \gamma| = n_a \nonumber \\
%%|\forall c(b) \in P_p| = n_b \nonumber
%%\end{eqnarray}

\section{Performance Evaluation}
\label{sec:performance_evaluation}

In this section, we discuss a quantitative analysis of the performance of ESPOON. It should be noticed that here we are concerned about quantifying the overhead introduced by the encryption operations performed both at the trusted environment and the outsourced environment. In the following discussion, we will not take into account the latency introduced by the network communication.

\subsection{Implementation Details}
We have implemented ESPOON in Java $1.6$. We have developed all the components of the architecture required for performing the policy deployment and policy evaluation phases. For the cryptographic operations, we have implemented all the functions presented in the previous section.

We have tested the implementation of ESPOON on a single node based on an Intel Core2 Duo $2.2$ GHz processor with $2$ GB of RAM, running Microsoft Windows XP Professional version $2002$ Service Pack $3$. The number of iterations performed for each of the following results is $1000$.

\subsection{Performance Evaluation of the Policy Deployment Phase}
In this section, we analyse the performance of the policy deployment phase. In this phase, the access policies are first encrypted at the Admin User side (that is a trusted domain) and then sent over to the Administration Point running on the outsourced environment. Here, the policies are re-encrypted and stored in the Policy Store on the outsourced environment.

Our policy representation consists of the tree representing the policy condition and the $\langle S, A, T \rangle$ tuple describing what action $A$ a subject $S$ can perform over the target $T$. In the condition tree, leaf nodes can represent string comparison (for instance, ``Location=HR-WARD'') and/or range comparison over numerical attributes (for instance, ``Access Time $>$ 9''). While a string comparison is always represented by a single leaf node in the condition tree and a single range comparison over numerical attributes may require more than one leaf node. In the worst case, a single range comparison on a value represented as a $s$-bit bag may require $s$ separate leaf nodes. Therefore, the range comparisons in a condition tree have a major impact on the encryption of a policy at deployment time.

\begin{figure*}[ht]
\centering
\subfigure[]{
% left bottom right top
\includegraphics[trim=20mm 20mm 45mm 70mm,clip,width=.38\textwidth]{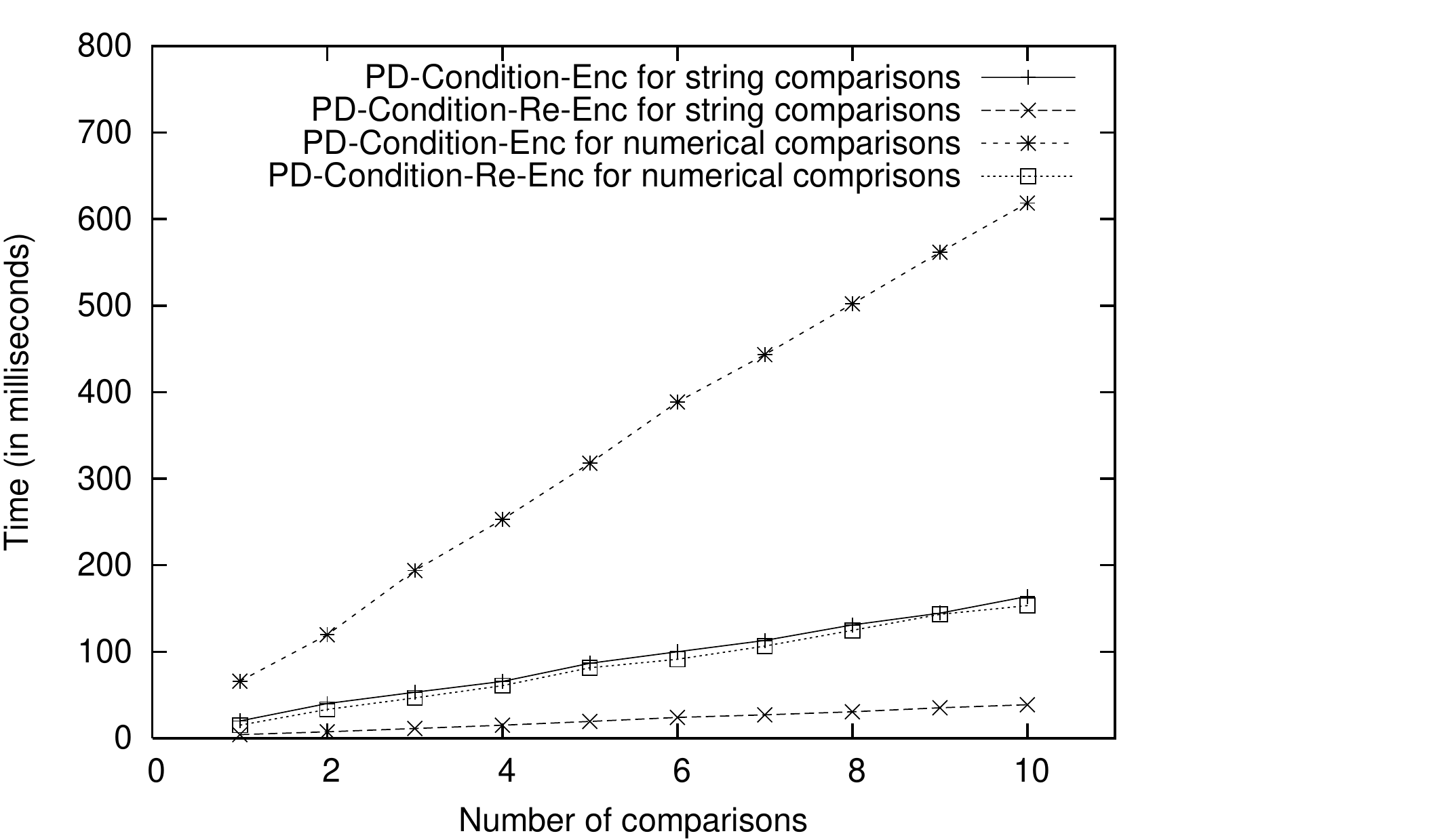}
\label{fig:espen-cc-attr-enc}
}
\subfigure[]{
% left bottom right top
\includegraphics[trim=20mm 20mm 140mm 110mm,clip,width=.3\textwidth]{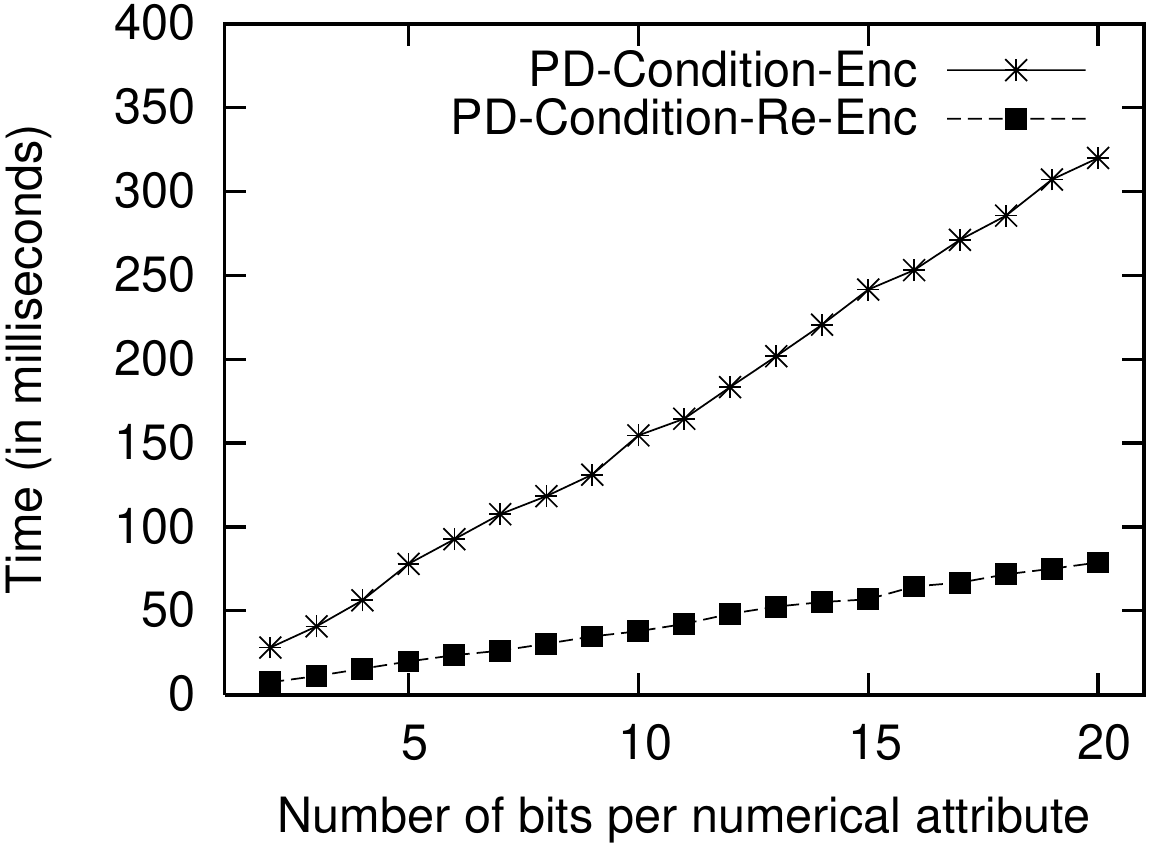}
\label{fig:espen-cc-num-bits-enc}
}
\label{fig:performance_deployment}
\caption{Performance overhead of access policy condition during the policy deployment}
%\caption[Performance overhead of access policy condition during the policy deployment]%{Caption of subfigures \subref{fig:subfig1}, \subref{fig:subfig2} and \subref{fig:subfig3}}
\end{figure*}

This behaviour is shown in Figure \ref{fig:espen-cc-attr-enc}. In the figure, we increase the number of string comparisons and numerical comparisons present in a condition tree of an access policy. As the graphs show, the time taken by $PD \mhyphen Condition \mhyphen Enc$ and $PD \mhyphen Condition \mhyphen Re \mhyphen Enc$ functions are growing linearly with the number of comparisons in the policy condition tree. However, the range comparisons over numerical attributes have a steeper line. For this testing, we have used the following comparisons. For the string we used ``$attributeName_i$=$attributeValue_i$'', where $i$ varies from 1 to 10. For the numerical comparison, we used ``$attributeName_i<15$\#4''.\footnote{It should be noted that using the comparison less than 15 in a 4 bits representation represents the worst case scenario requiring 4 leaf nodes.}

To check how the size of the bit representation impacts on the encryption functions during the deployment phase, we performed the following experiment. We fixed the number of numerical comparisons in the condition tree to only one and increased the size $s$ of the bit representation from $2$ to $20$ for the comparison ``$attributeName<2^s-1$. Figure \ref{fig:espen-cc-num-bits-enc} shows the performance overhead of the encryption during the policy deployment phase for the $PD \mhyphen Condition \mhyphen Enc$ and $PD \mhyphen Condition \mhyphen Re \mhyphen Enc$ functions. We can see that the policy deployment time incurred grows linearly with the increase in the size $s$ of a numerical attribute. In general, the time complexity of the encryption of the tree condition during the policy deployment phase is $O(m+ns)$ where $m$ is the number of string comparisons, $n$ is the number of numerical comparisons, and $s$ represents the number of bits in each numerical comparison.

\begin{table}[htp]
\centering
\caption{Performance overhead of encrypting the $\langle S, A, T \rangle$ tuple during the policy deployment}
\label{tab:sat_pol_deployment}
{\small
\begin{tabular}{ |l|c|c| }
\hline
\textbf{Function Name} & \textbf{PD-SAT-Enc} & \textbf{PD-SAT-Re-Enc} \\ \hline
Time (in milliseconds) & 46.44 & 11.65 \\ \hline
\end{tabular}
}
\end{table}

As for the $\langle S, A, T \rangle$ tuple, the average encryption time taken by the $PD \mhyphen SAT \mhyphen Enc$ and $PD \mhyphen SAT \mhyphen Re \mhyphen Enc$ are shown in Table \ref{tab:sat_pol_deployment}. The time complexity of the encryption of the $\langle S, A, T \rangle$ tuple during the policy deployment phase does not depend on any parameters and can be considered constant.

During the policy deployment phase, the encryption operations performed at the Admin User side take more time to encrypt the access policy than the SP side to re-encrypt the same policy (either $PD \mhyphen Condition \mhyphen Re \mhyphen Enc$ or $PD \mhyphen SAT \mhyphen Re \mhyphen Enc$). This is because the $PD \mhyphen Condition \mhyphen Enc$ or $PD \mhyphen SAT \mhyphen Enc$ functions perform more complex cryptographic operations, like generation of random number and hash calculations, than the respective functions on the SP side.

\begin{figure*}[ht]
\centering
\subfigure[]{
%% left bottom right top
\includegraphics[trim=20mm 20mm 140mm 110mm,clip,width=.3\textwidth]{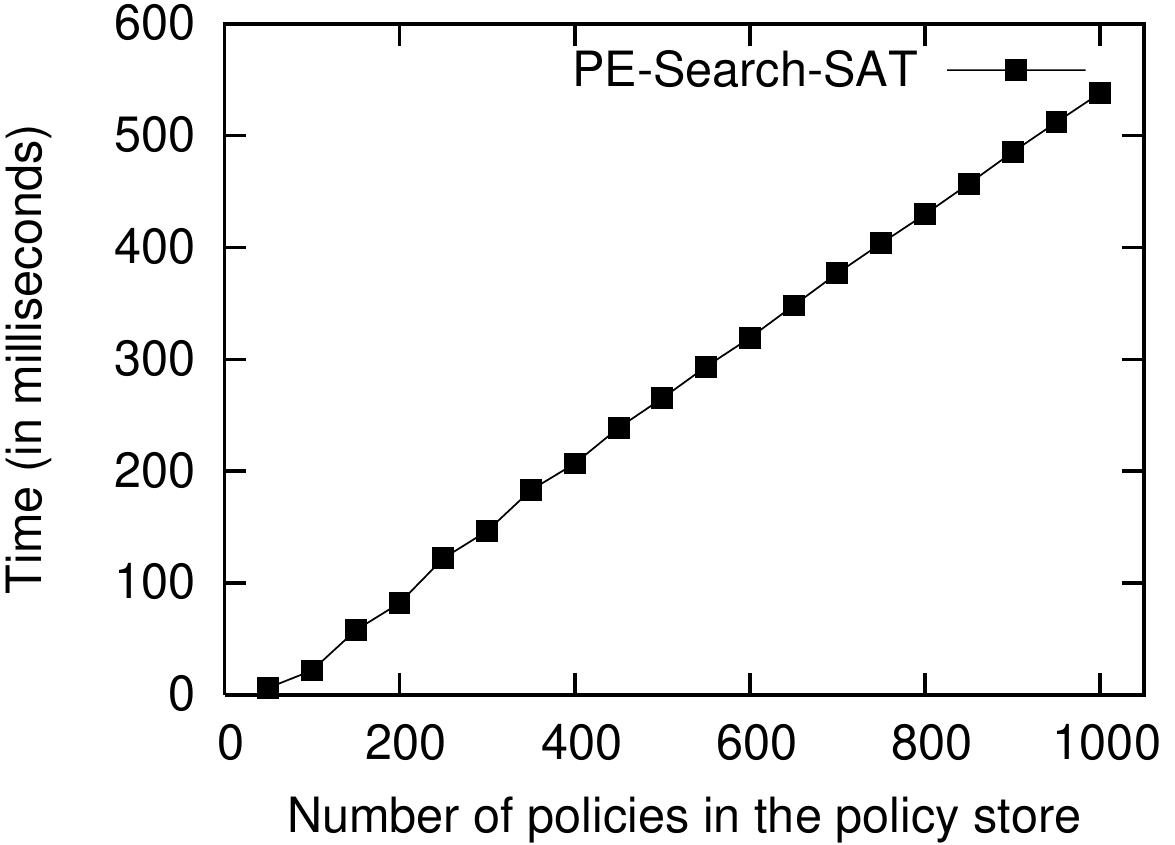}
\label{fig:espen-sat-search}
}
\subfigure[]{
% left bottom right top
\includegraphics[trim=20mm 20mm 115mm 90mm,clip,width=.31\textwidth]{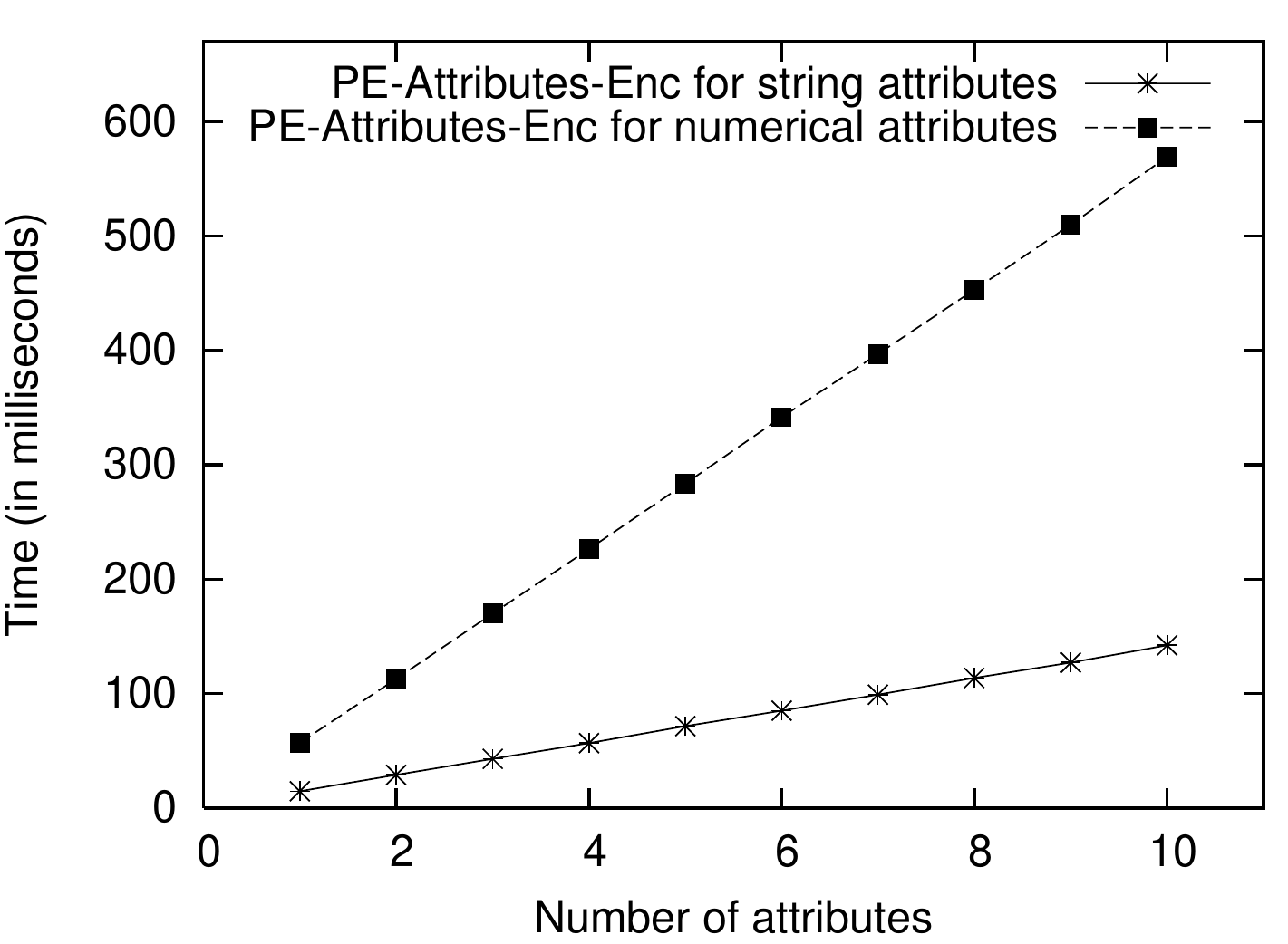}
\label{fig:espen-requester-enc}
}
\subfigure[]{
% left bottom right top
\includegraphics[trim=20mm 20mm 90mm 75mm,clip,width=.32\textwidth]{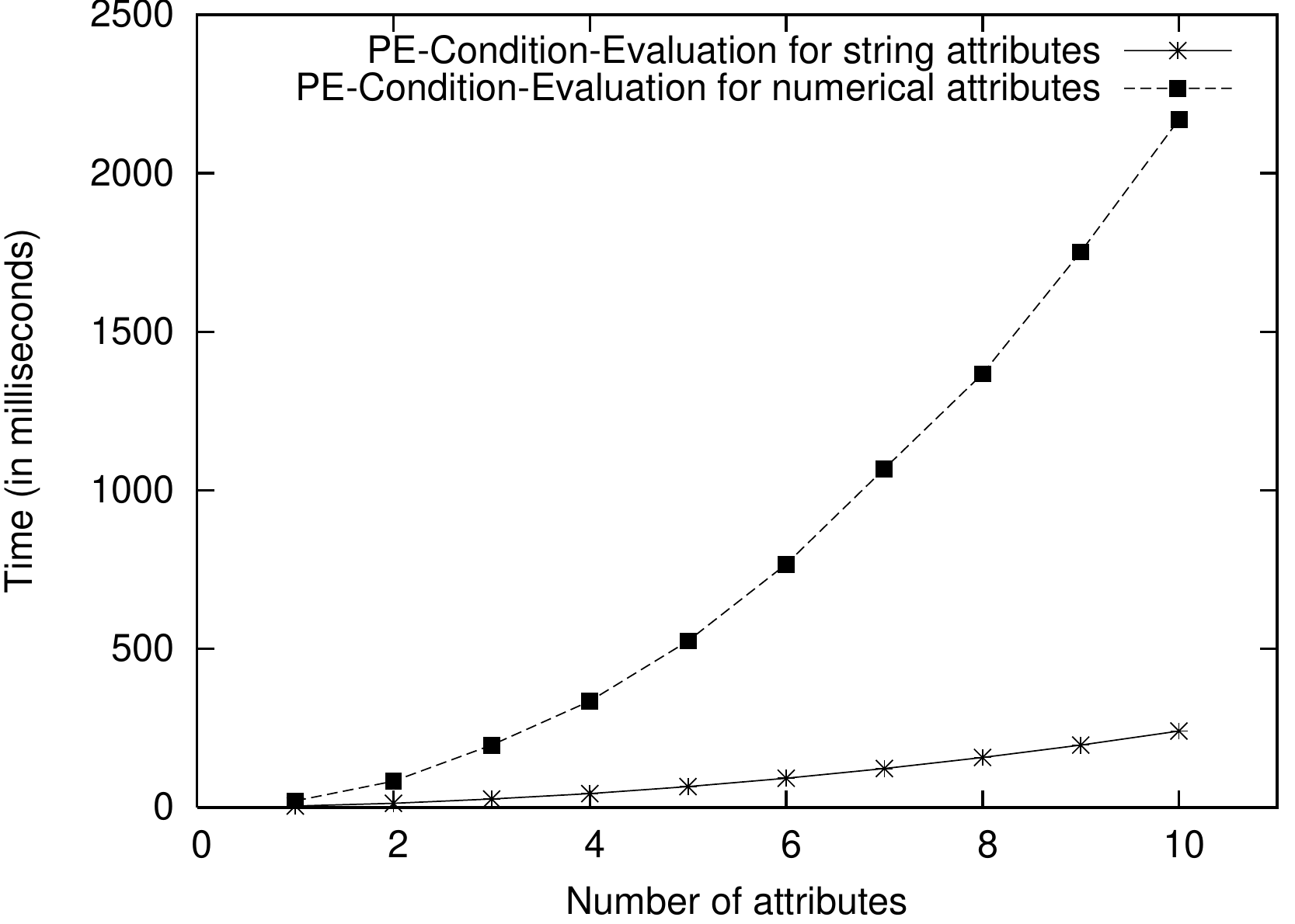}
\label{fig:espen-server-match}
}
\label{fig:performance_evaluation}
\caption{Performance overhead of access policy condition during the policy evaluation}
%\caption[Performance overhead of access policy condition during the policy deployment]%{Caption of subfigures \subref{fig:subfig1}, \subref{fig:subfig2} and \subref{fig:subfig3}}
\end{figure*}

\subsection{Performance Evaluation of the Policy Evaluation Phase}
In this section, we analyse the performance of the policy evaluation phase, where a Requester sends the encrypted request to the PEP running on the outsourced environment. The PEP forwards the encrypted request to the PDP that has to select the set of policies that are applicable to the requests. Once the PDP has found the policies then the PDP will evaluate if the attributes in the contextual information satisfy any of the conditions of the selected policies.

To make a request, it is necessary to generate the $\langle S, A, T \rangle$ tuple representing the subject $S$ requesting to perform action $A$ on target $T$. The $\langle S, A, T \rangle$ tuple needs to be transformed into \emph{trapdoors} before they are sent over to the PEP on the outsourced environments. The trapdoors will be used for performing the encrypted policy evaluation on the outsourced environment. The trapdoor representation does not leak information on the element of the $\langle S, A, T \rangle$ tuple.

Once the PDP gets the request as a list of trapdoors, the PDP performs an encrypted search in the Policy Store looking for any stored policies with matching $\langle S, A, T \rangle$ tuples. Figure \ref{fig:espen-sat-search} shows the performance overhead incurred by the encrypted search. In our testing, we varied the number of encrypted policies stored in the Policy Store ranging from $50$ to $1000$. As we can observe, it takes $0.5$ milliseconds on average for performing an encrypted match operation between the $\langle S, A, T \rangle$ tuple of the request and the $\langle S, A, T \rangle$ tuple of a stored policy. This means that on average it takes half a second for finding a matching policy in the Policy Store with $1000$ policies.

If any match is found in the Policy Store then the PDP needs to fetch the contextual information from the PIP. The PIP is responsible to collect and send the required contextual information that might be containing environmental and Requester attributes. The PIP transforms these attributes into trapdoors before they sent over to the PDP. Once again, the trapdoor representation does not leak information on the actual value of the attribute. Figure \ref{fig:espen-requester-enc} shows the performance of generating trapdoors by the PIP. The performance of function $PE \mhyphen Attributes$ in Figure \ref{fig:espen-requester-enc} is almost similar to the function $PD \mhyphen Condition \mhyphen Enc$ called by the Admin User during the policy deployment phase as shown in Figure \ref{fig:espen-cc-attr-enc}.

When the policies with matching $\langle S, A, T \rangle$ tuples have been found, then the PDP will perform the evaluation of the condition tree of the policies. To evaluate the tree condition, the PDP matches the Requester and environmental attributes against the leaf nodes in the condition tree using $PE \mhyphen Condition \mhyphen Evaluation$ function. To quantify the performance overhead of the encrypted matching of the Requester and environmental attributes against the condition tree we have performed the following test. First of all, we have considered two cases: the first case is the one in which only string attributes are provided by the PIP and the tree condition contains only string comparisons; in the second case only numerical attributes are provided by the PIP and the tree condition consists only of numerical comparisons. For both cases, the number of attributes varies together with the number of comparisons in the tree. In particular, if the request contains $n$ different attributes then the tree condition will contain $n$ different comparisons.

Figure \ref{fig:espen-server-match} shows the results for both cases. As we can see, the condition evaluation for the numerical attributes has a steeper curve. This can be explained as follows. For the first case, for each string attribute only a single trapdoor is generated. A string comparison is represented as a single leaf node in the condition tree. This means that $n$ trapdoors in a request are matched against $m$ leaf nodes in the tree resulting in a $O(nm)$ complexity (however, in our experiments the number of attributes and the number of comparisons are always the same). For the case of the numerical attributes, we have also to take in to consideration the bit representation. In particular, for a give numerical attribute represented as $s$ bits, we need to generate $s$ different trapdoors. This means that $n$ numerical attributes in a request will be converted in to $n s$ different trapdoors. These trapdoors then need to be matched against the leaf nodes representing the numerical comparisons. As we have discussed for the policy deployment phase, in the worst case scenario, a numerical comparison for a $s$-bit numerical attribute requires $s$ different leaf nodes. In a tree with $m$ different numerical comparisons, this means that the $n s$ trapdoors need to be matched against $m s$ resulting in $O(n m s^2)$ complexity.

\section{Conclusions and Future Work}
\label{sec:conclusion_future_work}

In this paper, we have presented the ESPOON architecture to support policy-based access control mechanism for outsourced and untrusted environments. Our approach separates the security policies from the actual enforcing mechanism while guaranteeing the confidentiality of the policies when given assumptions hold (i.e., the SP is honest-but-curious). The main advantage of our approach is that policies are encrypted but it still allows the PDP to perform the policy evaluation while not learning anything on the policies. Second, ESPOON is capable of handling complex policies involving non-monotonic boolean expressions and range queries. Finally, the authorised users do not share any encryption keys making the process of key management very scalable. Even if a user key is deleted or revoked, the other entities are still able to perform their operations without requiring re-encryption of the policies.

%TODO: Mention about provenance as a future work, cite RSA paper on impossibility of encryption alone...

As future directions of our research, we are working on integrating a secure audit mechanism in ESPOON. The mechanism should allow the SP to generate genuine audit logs without allowing the SP to get information about both the data and the policies. However, an auditing authority must be able to retrieve information about who accessed the data and what policy was enforced for any access request made. Another direction of our work is towards the extension of the encrypted search and match capabilities to handle the case of negative authorisation policies and policies for long-lived sessions where the conditions need to be continuously monitored and the attributes of the request can be dynamically updated.

% TODO: auto collection of attributes from PIP

% ... The primary challenge in this line of work is to find a
%what is done, what needs to be done and open problems.
%% Fault tolerant policy based encryption.
%
%Once the government or audit (log) authority notice that a doctor has misused the data then they may deny the doctor's access by revoking her license (technically, her key will be revoked). For revocation, the trusted KGC removes the keypair and reports to CSP keys store. Therefore, a CSP keys store is always checked first when a user makes a request to access data.

%policy cannot be separated from the data
%CSP performs right actions
% TODO: Audit log (secure & integral)
% - proof system if a resource is accessed

% TODO: PEP as an open issue

% TODO: forensic in cloud architecture: audit trails: Audit log corruption in cloud is an open problem

% TODO: how to verify if policy is enforced

% TODO: type and identity based encryption: refer to UT paper by TA

% TODO: consider PDP in distributed environment (future work?)

% TODO: who generates contextual attributes.

% Bottle neck: how to find/calculate automatic generation of trapdoors

% Minimal disclosure

% TODO: trusted KGC can work as an Ephemerizer

% TODO: how to withstand DoS attack on the storage server.

% TODO: add acknowledgement for ENDORSE


\section*{Acknowledgment}
The work of the first and third authors is supported by the EU FP7 programme, Research Grant 257063 (project ENDORSE).

\bibliographystyle{IEEEtran}
\bibliography{IEEEabrv,enc-policy}

\end{document}